\documentclass[preprint,12pt]{elsarticle}

\usepackage{lineno}

\usepackage{amssymb}
\usepackage{epstopdf}

\journal{Nuclear Instruments and Methods in Physics Research A}

\begin{document}

\begin{frontmatter}



\title{A Generalized Model for Light Transport in Scintillators}


\author[label1,label2]{Jiangshan Lan$^{\dag}$}
\ead{jiangshanlan@impcas.ac.cn}
\author[label1,label2]{Meng Ding}
\author[label1,label2]{Chengdong Han}
\author[label1]{Yapeng Zhang}
\author[label1]{Xurong Chen$^{\ddag}$}
\ead{xchen@impcas.ac.cn}
\address[label1]{Institute of Modern Physics, Chinese Academy of Sciences, Lanzhou 730000, China}
\address[label2]{University of Chinese Academy of Sciences, Beijing 100049, China}

\begin{abstract}
Transported light in the medium usually shows as an exponential decay tendency. In the DAMPE strip scintillators, however, the phenomenon of light attenuation as the hit position approaches the end of the scintillator can not be described by the simple exponential decay model.
The spread angle of PMT relative to hit position is distance-dependent, so the larger the angle, the larger the proportion $\eta$ of emitted light to becomes the effective input light. We consider the contribution of the spread angle, and propose a generalized model: $f(x)=I_0\eta_{a}(x)e^{-x/\lambda}+I_0\eta_{b}(x)e^{-(2L-x)/\lambda }$.
The model well describes the light attenuation in the scintillator, reducing the maximum deviation of the sample from the fit function from 29\% to below 2\%. Moreover, our model contains most of the traditional models, so the experimental data that traditional models can fit and our models fit well.
\end{abstract}

\end{frontmatter}

\section{Introduction}\label{Introduction}

Scintillation detector, as an important detector, is widely used in time-of-flight detector, position detector and electromagnetic calorimeter, including CMS at CERN\cite{LHC}, BESIII at HIEP\cite{BESIII}, PSD and BGO at DAMPE\cite{DAMPE,zhou,Yu2017,YpZhang2017} \emph{et al}. When particles enter the scintillator, the incident particle deposits part energy or all the energy, which causes ionization and excitation in a scintillator and emits fluorescence of a certain wave length. Then the emitted fluorescence propagates to the end of the scintillator and is collected by photomultiplier tube(PMT). Due to the particular importance of light transmission, various models have been tried to account for the decay of scintillators in the past few decades. It is worth mentioning that the double exponent(DE) model\cite{Kaiser64} and the reflected back(RB) model\cite{Taiuti96} are now generally accepted. Unfortunately, none of them well explain the anomalous enhancement of the light signals  along the scintillator ends in DAMPE.

If the light transport can be clearly analyzed in a scintillator, we can accurately determine two important parameters -- hit position and deposited energy. In this paper, because the spread angle of PMT relative to hit position is distance-dependent, we propose a generalized model of light transport in the scintillator. The model well describes the light attenuation in the scintillator, and is used to fit samples from PSD at DAPME \cite{zhou,Yu2017,YpZhang2017}, and gives the explanation of the phenomenon of samples different from the fitting function along the scintillator ends, which has not been explained well for many years.

The organization of this paper is as follows. A brief introduction about the exponential decay(ED) model, the double exponent(DE) model and the reflected back(RB) model in Sec.\ref{Review}. Sec.\ref{NewModel} considers the contribution of the spread angle and gives our model about light transport in a scintillator. Sec.\ref{Results} gives the result of our model. Sec.\ref{Discussion} explains why is our new model so powerful. Finally, conclusion are given in Sec.\ref{Conclusion}.

\section{Traditional Models for Light Transport in Scintillators}\label{Review}

The striped and the parallelepiped scintillators are the most widely used scintillators, including the calorimeter detector, time-of-flight detector and particle position detector. When particles hit the scintillator, the particle energy deposits in the scintillator. Then the atomic in the scintillator transition to the excited state, and back after excitation fluorescence, the emitted light is collected by the PMT at the end of the scintillator. Thus, the information of particles can be extracted by analyzing electronic information. It is generally known that transmitted light of an initial intensity $I_0$ in a medium can be described by an exponential decay(ED) function with a $\lambda$ attenuation length:

$$I(x)=I_{0}e^{-x/\lambda }.$$

In the traditional model, the striped scintillator is assumed to be a line of length $L$. When the particle hits on the scintillator away from the PMT to the $x$ position, the collected light can be express as
\begin{equation}\label{Eq.exp}
f(x)=I_{0}e^{-x/\lambda }.
\end{equation}
And the deviation between fitting function and the sample can be described as
\begin{equation}\label{Eq.dev}
\delta(x)=\frac{f_{samples}(x)-f_{fitting}(x)}{f_{fitting}(x)}.
\end{equation}
This ED model was used to fit samples of PSD on DAMPE \cite{zhou,Yu2017,YpZhang2017}, and showed a sharp increase in deviation as the hit position approaches the extremity of the scintillator. Therefore we searched for further physical models to fit the samples. In the past several decades, there are two further physical models: the Double Exponential(DE) model\cite{Kaiser64} and the Reflected Back(RB) model\cite{Taiuti96}.

\begin{figure}
\begin{center}
\includegraphics[width = 0.43\textwidth]{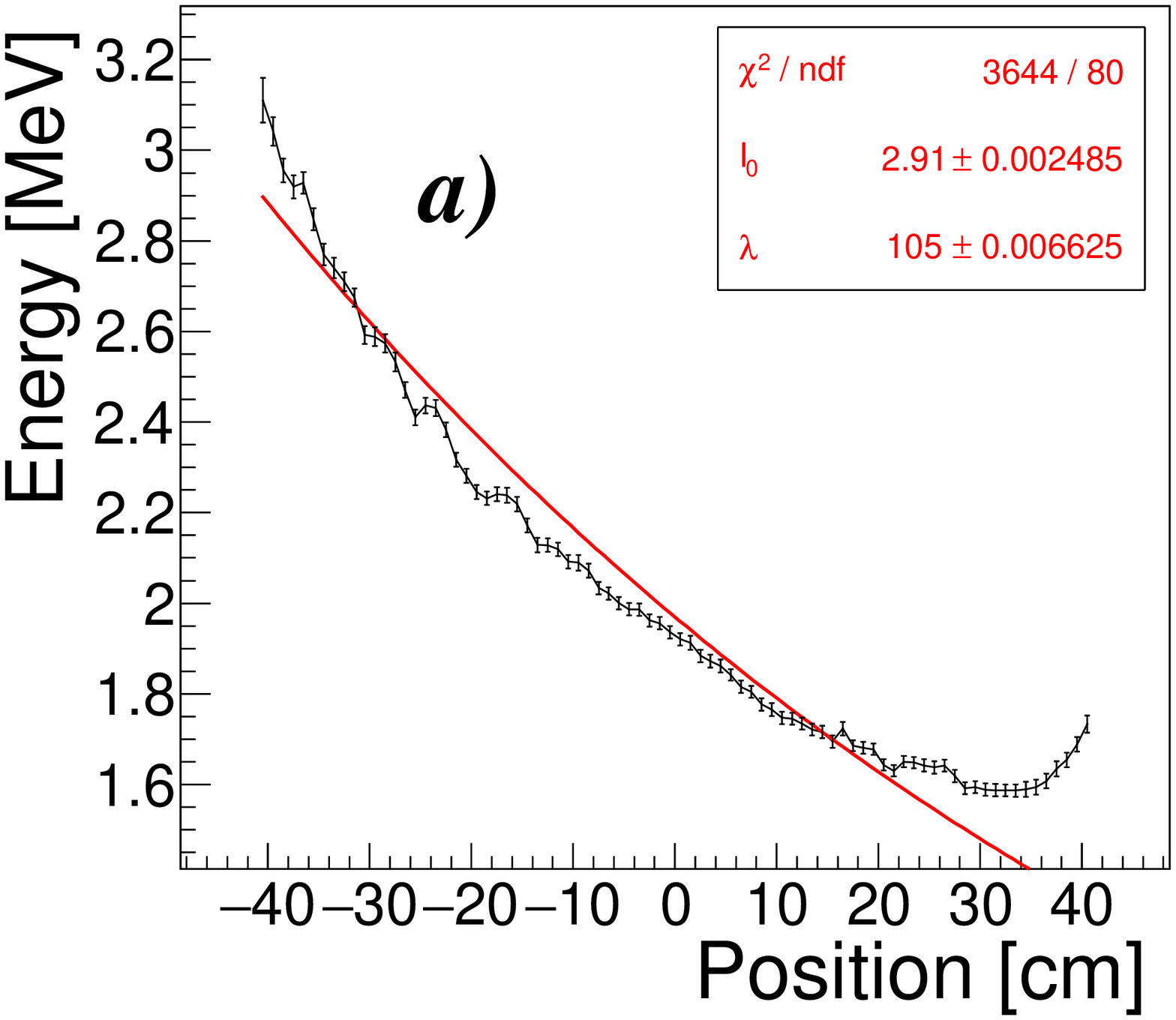}
\includegraphics[width = 0.43\textwidth]{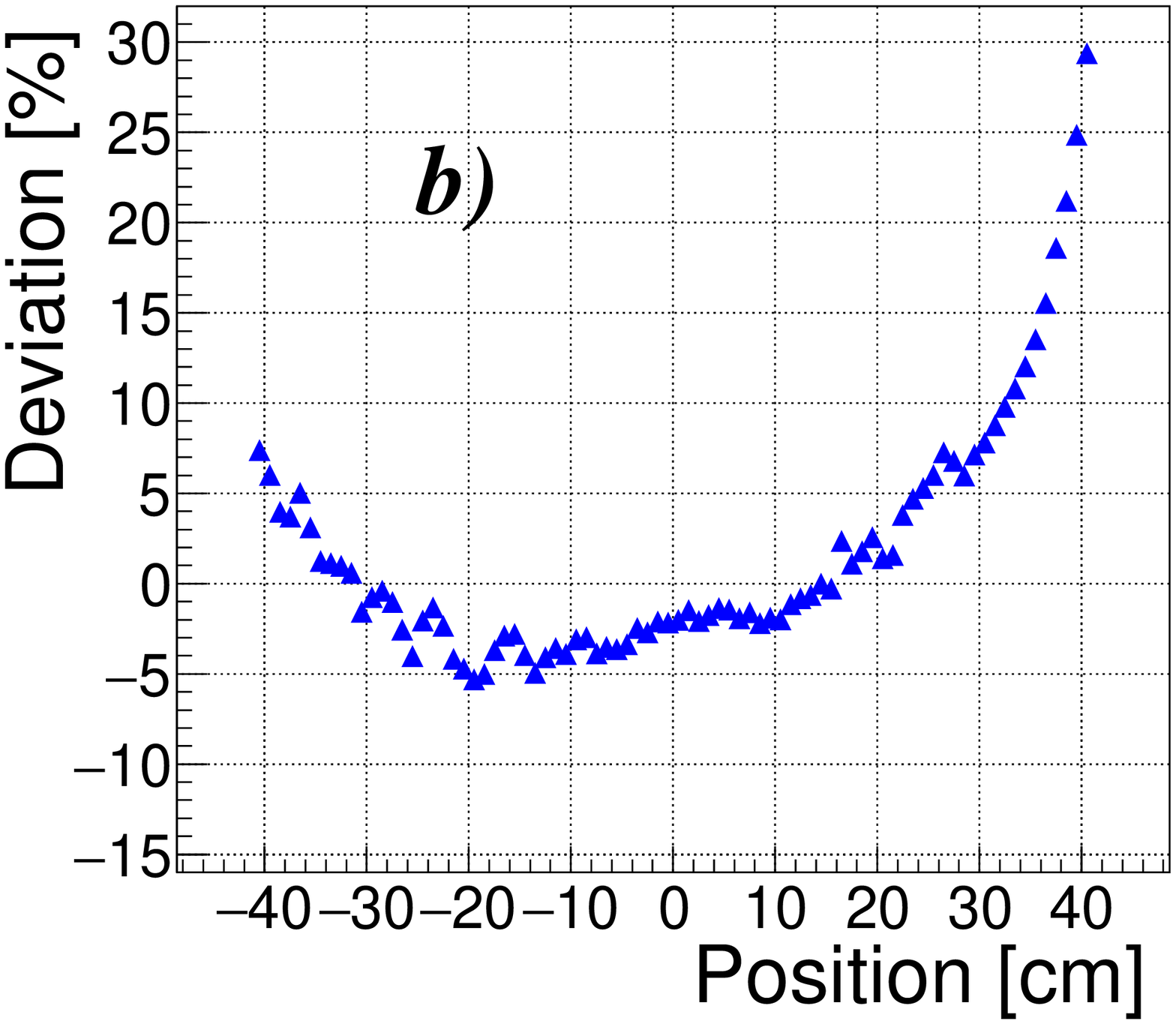}
\includegraphics[width = 0.43\textwidth]{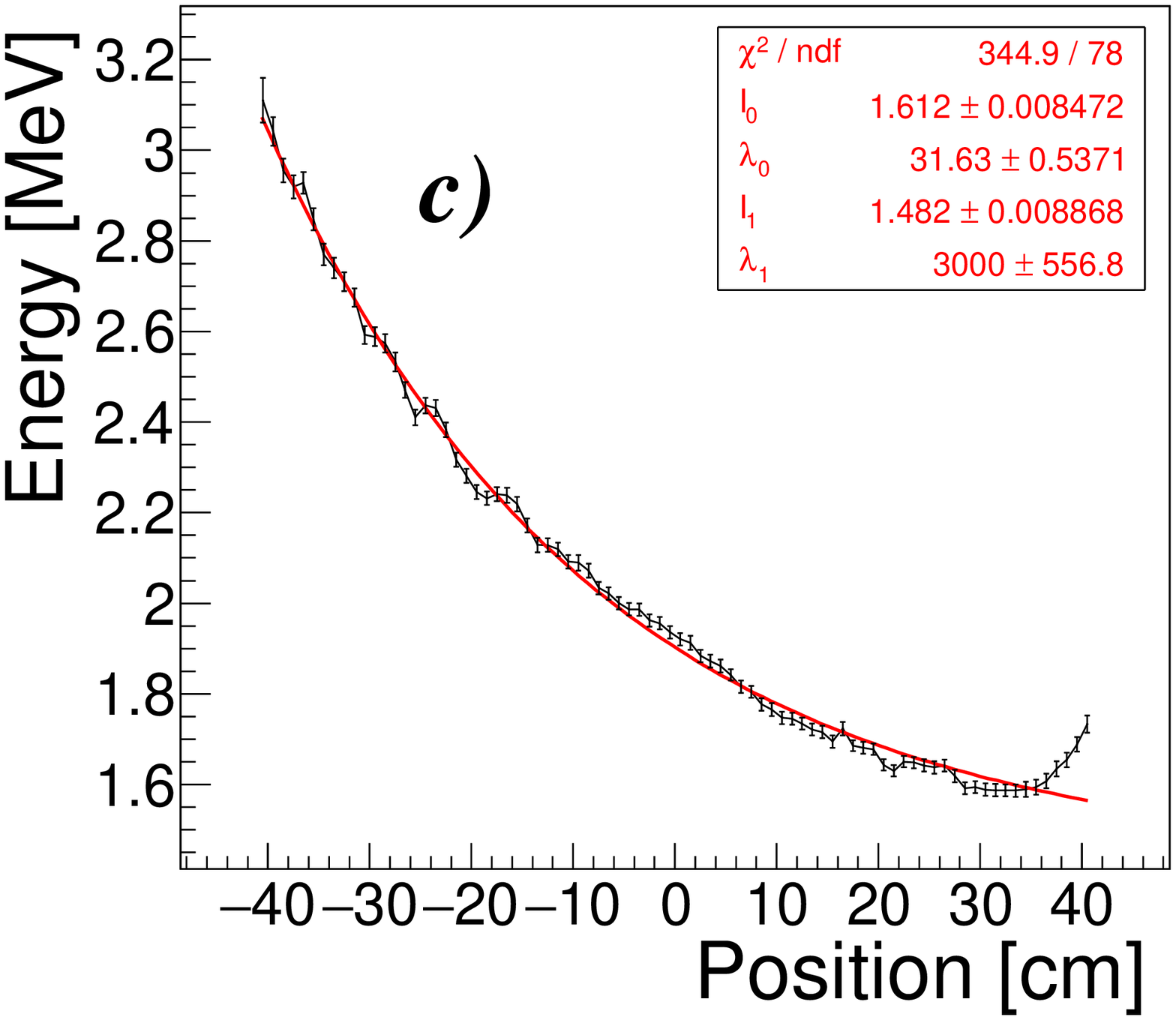}
\includegraphics[width = 0.43\textwidth]{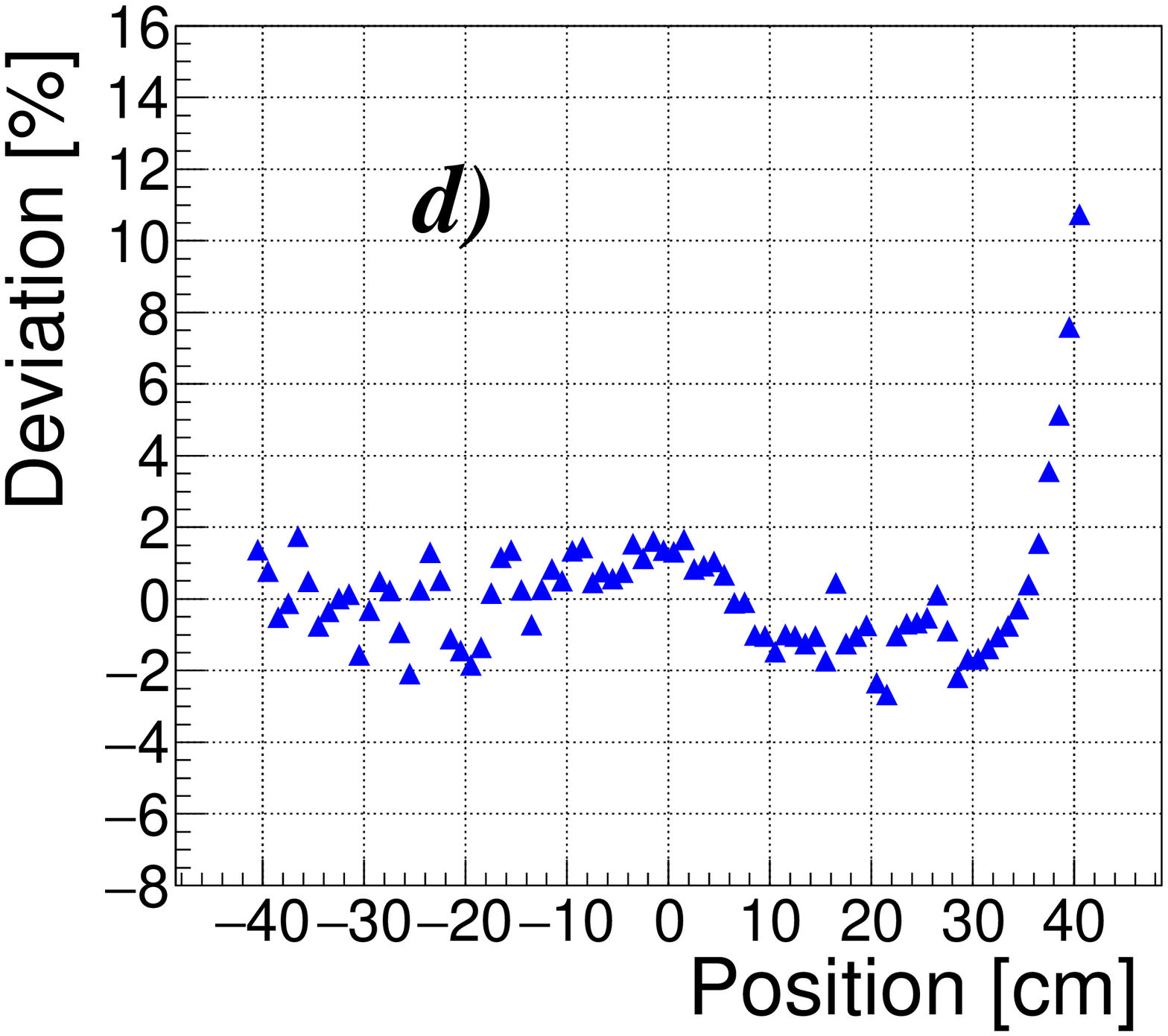}
\includegraphics[width = 0.43\textwidth]{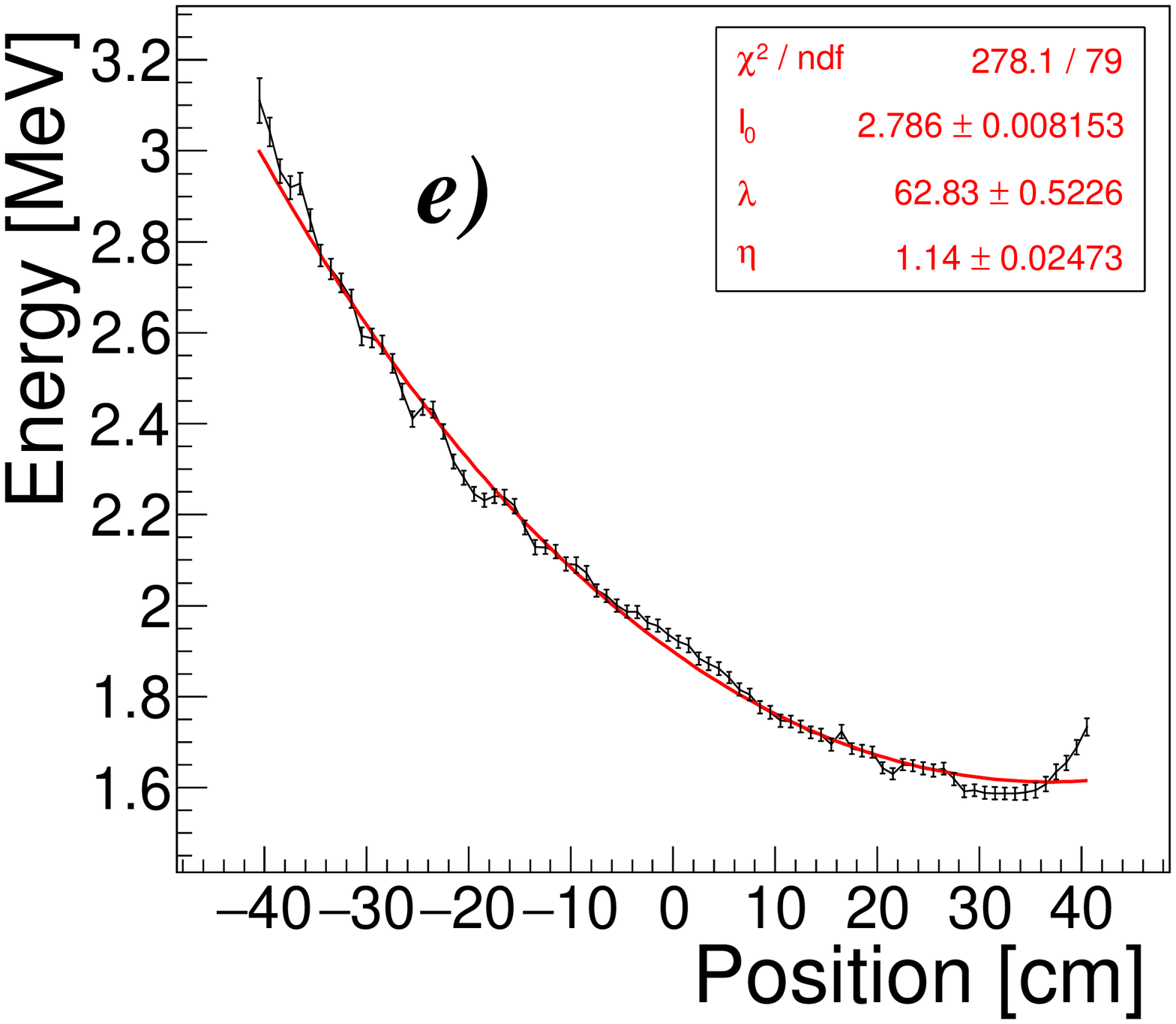}
\includegraphics[width = 0.43\textwidth]{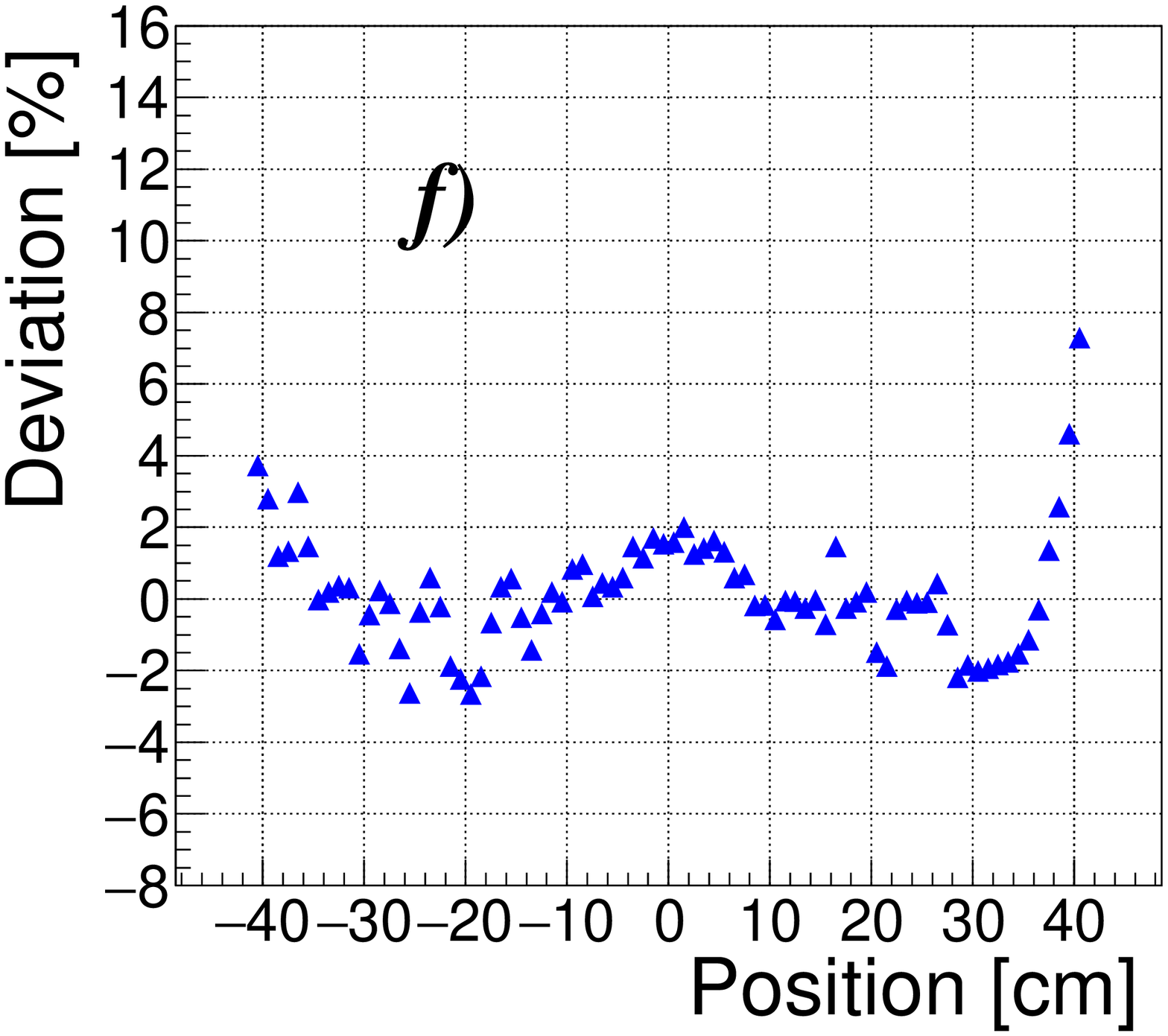}
\caption{a) The curve fitted with the samples from PSD at DAMPE \cite{zhou,Yu2017,YpZhang2017} by the Exponential Decay model.
 b) The deviation between the Exponential Decay model fitting function and the samples.
 c) The curve fitted with the same samples by the Double Exponent model.
 d) The deviation between the Double Exponent model fitting function and the samples.
 e) The curve fitted with the smae samples by the Reflected Back model.
 f) The deviation between the Reflected Back model fitting function and the samples.}
\label{TraModel}
\end{center}
\end{figure}

Kaiser \emph{et al} proposed the DE model of light transmission with a short $\lambda_0$ and a long $\lambda_1$ attenuation length components described by the expression\cite{Kaiser64}
\begin{equation}\label{Eq.Kaiser}
f(x)=I_{0}e^{-x/\lambda_0}+I_{1}e^{-x/\lambda_1},
\end{equation}
with $\lambda_0$ being of the order of a few centimetres and  $\lambda_1$ ranging from 1 to several metres.
Taiuti \emph{et al} proposed the RB model that a fraction $\eta$ of the emitted light could be reflected back to the photomultiplier adding to the direct light a second component described by the expression\cite{Taiuti96}
\begin{equation}\label{Eq.Taiuti}
\label{CosmicFlowCal}
f(x)=I_{0}(e^{-x/\lambda}+\eta e^{-(2L-x)/\lambda}).
\end{equation}
Both the DE model and the RB model had upgraded the ED model, and their deviations had been much reduced relative to the ED model. However, Fig.\ref{TraModel} shows that there is a deviation of about $10\%$ for both models. In order to fit the samples, we consider the contribution of the spread angle and propose a generalized model in next section.

\section{A Generalized Model for Light Transport in Scintillators}\label{NewModel}
The optical signal intensity collected by PMT is related to the effective input light and the propagation loss. In this section, we point out that the effective light is $x$-dependent. Where $x$ is the distance between hit position and the PMT.

When particles hit a scintillator, the spread angle of PMT relative to hit position is $x$-dependent. Since the emitted light is $4\pi$ uniformly distributed, the larger the spread angle, the larger the proportion $\eta$ of the emitted light becomes effective input light. Therefore, the contribution of the spread angle should be considered when describing the propagation of light in scintillators.

In traditional ED model, since the spread angle decreases as $x$ increases,
we can simply multiply a fraction  $\eta_a(x)=e^{\frac{c_1}{c_2+x}}$ in place of the contribution of the spread angle.
Where $c_1$ is less than $c_2$.

If the emitted light may reflect back to the PMT as the RB model,
the contribution of the spread angle should also be considered.
Here the spread angle decreases as $L-x$ increases,
so we can also simply multiply a fraction $\eta_b(x)=\frac{c_3}{1-x/c_4}$to replace the contribution of the spread angle.
Where $c_3 $ belongs to $ [0,1]$ , $c_4$ is greater than $L$ and $L$ is the length of the scintillator.

So we can describe light propagation in scintillators as
\begin{equation}\label{Eq.our}
f(x)=I_0\eta_{a}(x)e^{-x/\lambda}+I_0\eta_{b}(x)e^{-(2L-x)/\lambda },
\end{equation}
where $\eta_a(x)$ and $\eta_b(x)$ represent the contribution from the spread angle.

\section{Results}\label{Results}
\subsection{Fitting The Same Samples } 
\begin{figure}
\begin{center}
\includegraphics[width = 0.43\textwidth]{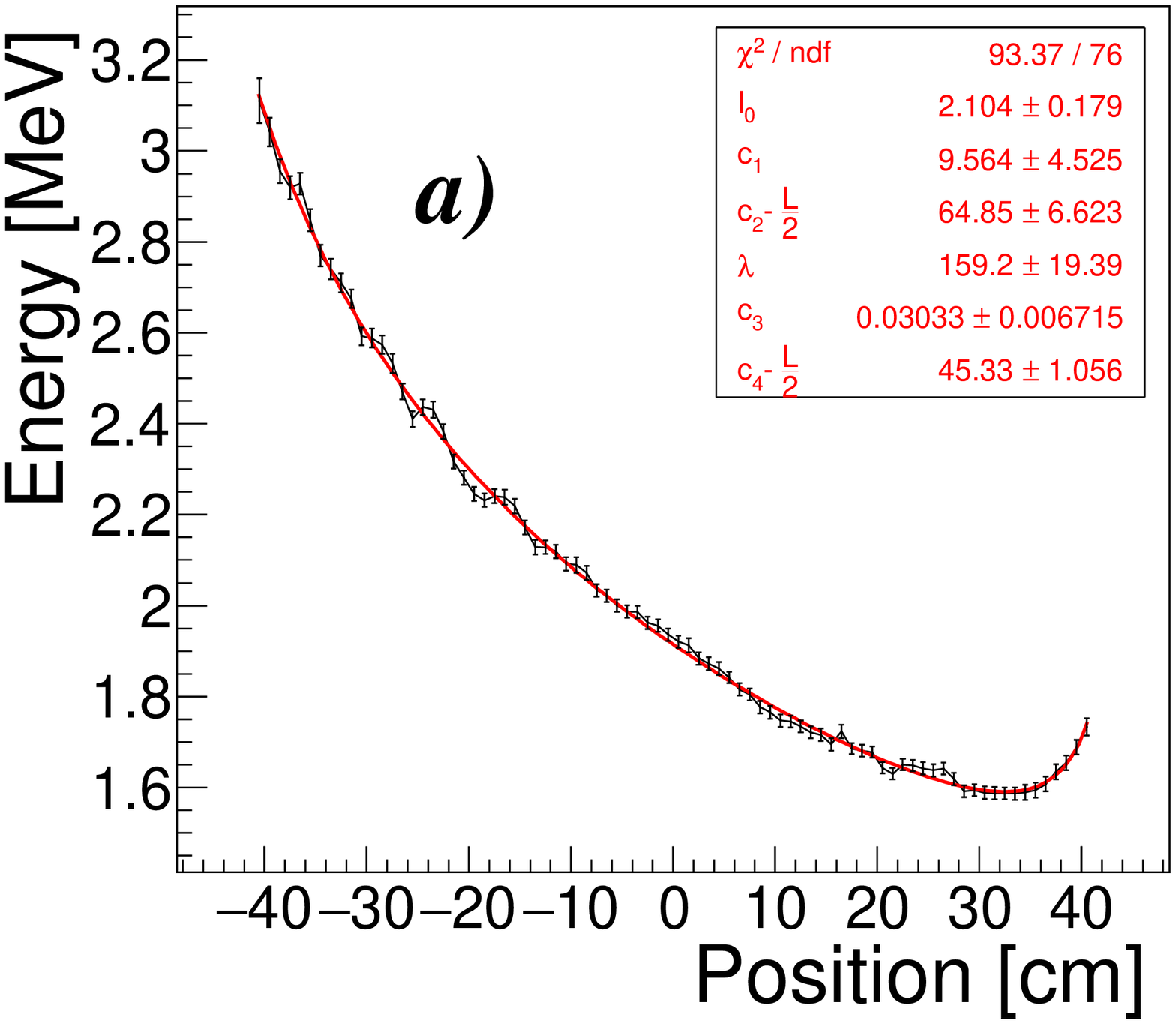}
\includegraphics[width = 0.43\textwidth]{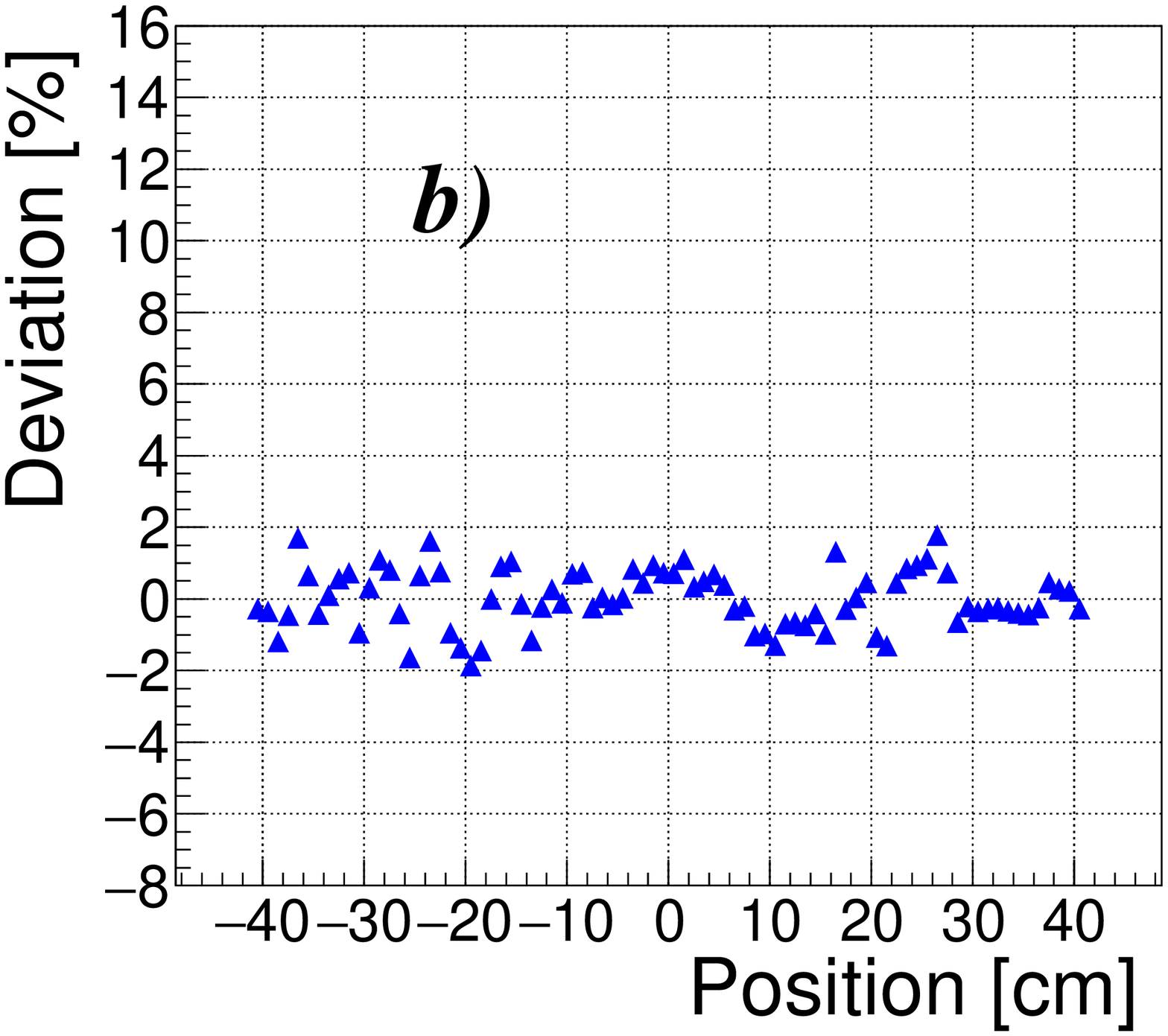}
\caption{a) The curve fitted with the samples from PSD@DAMPE by our new model.
 b) The deviation between our new model fitting function and the samples.}
\label{figDAMPE.our}
\end{center}
\end{figure}

Fitting the same samples with our model,
the deviation is reduced to less than 2\%.
Fig.\ref{figDAMPE.our} and Table.1 show that this is far better than 29\% of the ED model, 11\% of the DE model and 7\% of the RB model.

\begin{center}
Table.1: Compare The Deviation With Other Models
\begin{tabular}{ccccc}
	\hline &ED model&DE model&RB model&our model\\
	\hline 	$\delta_{max}$&29\%&11\%&7\%& $<$2\%\\
$\chi^{2}/ndf$&3644/80&344.9/78&278.1/79&93.4/76\\
	\hline
\end{tabular}
\end{center}
\subsection{More Experimental Data Fitting} 
\begin{figure}
\begin{center}
\includegraphics[width = 0.43\textwidth]{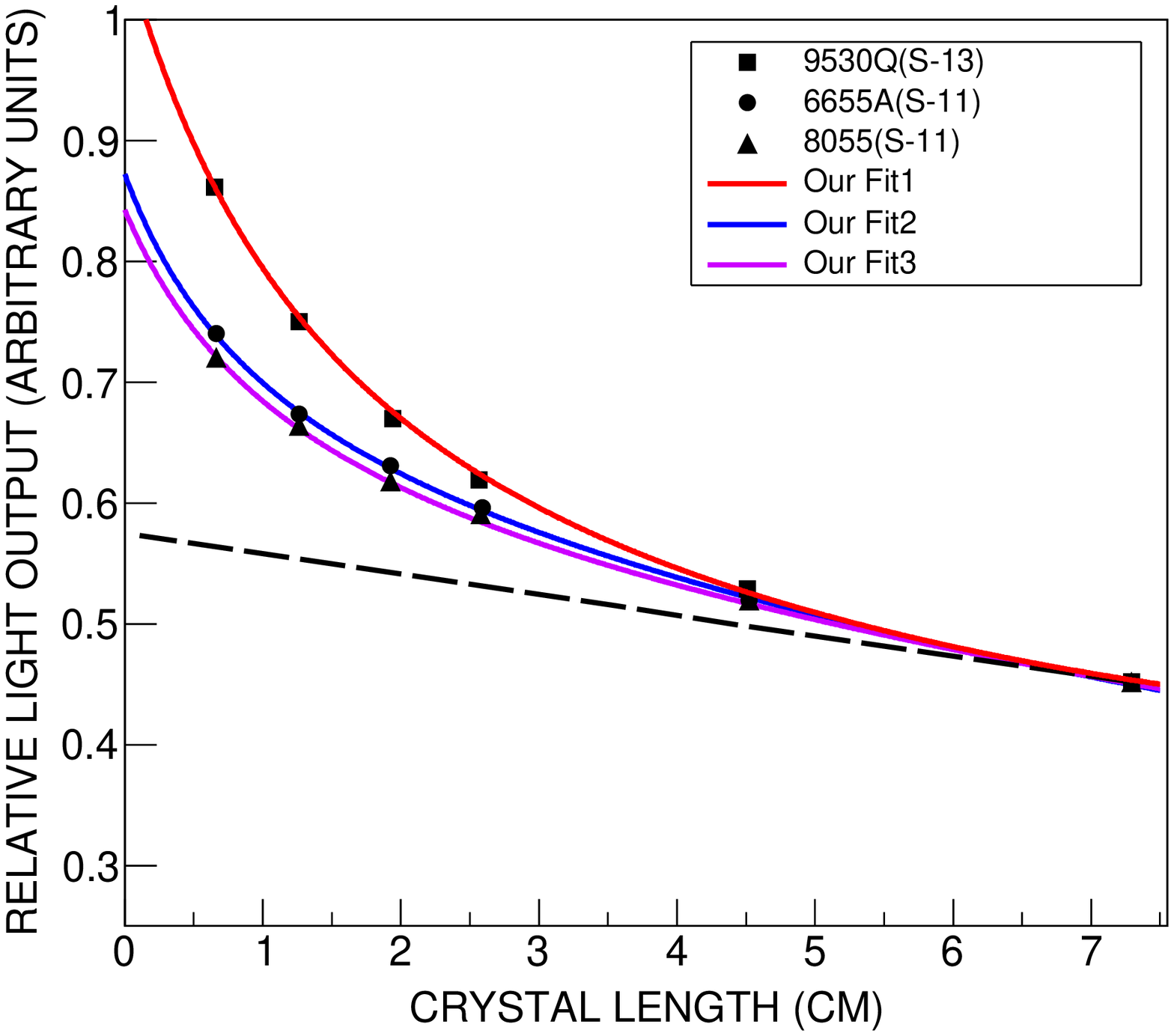}
\includegraphics[width = 0.43\textwidth]{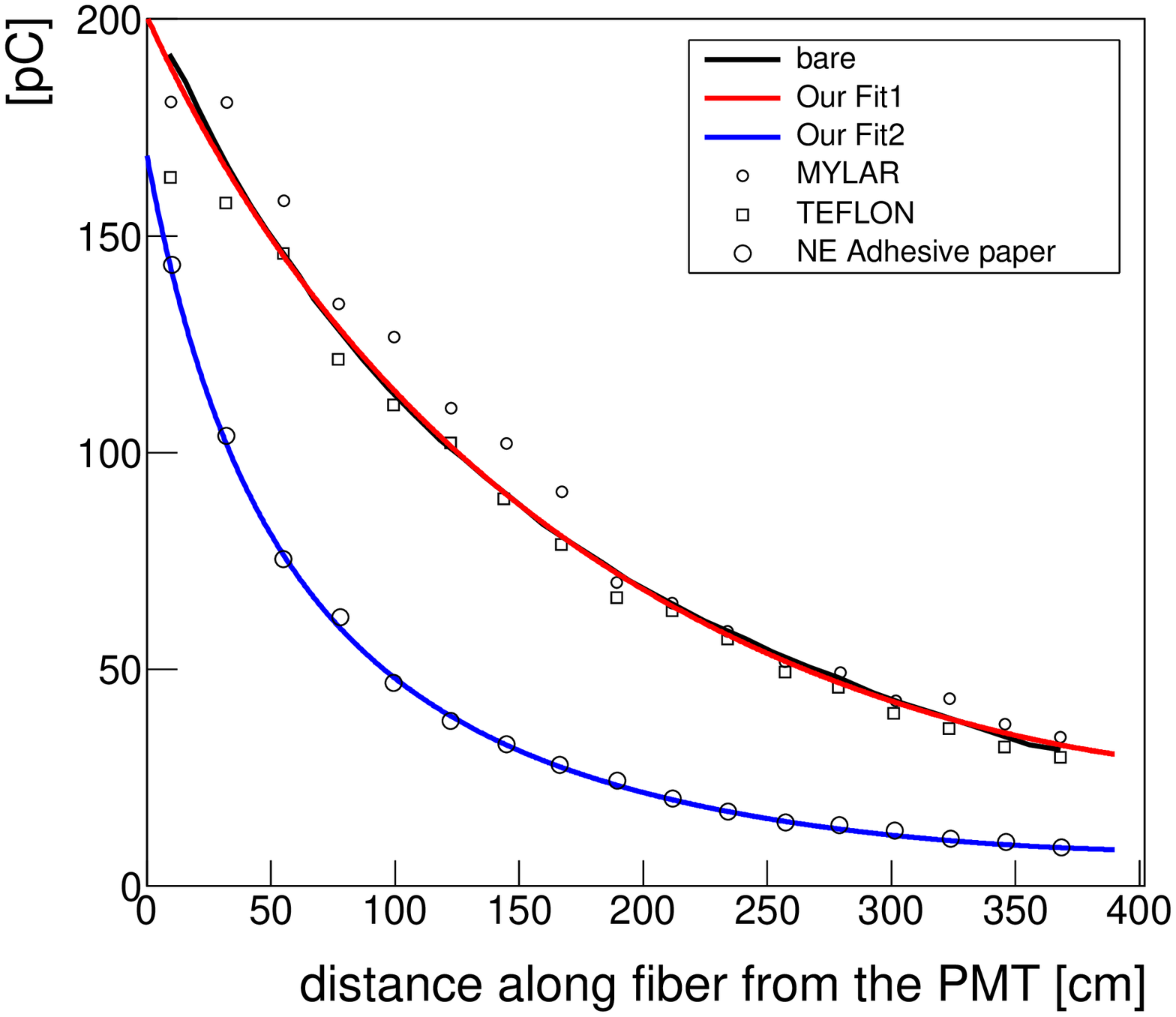}
\caption{left) Pilot B scintillator light output vs length using different PMT\cite{Kaiser64};
 right) the effect of wrapping material on the light transmission for different materials\cite{Taiuti96}. Our model fit show as line with color. }
\label{fig1.our}
\end{center}
\end{figure}

\begin{figure}
\begin{center}
\includegraphics[width = 0.43\textwidth]{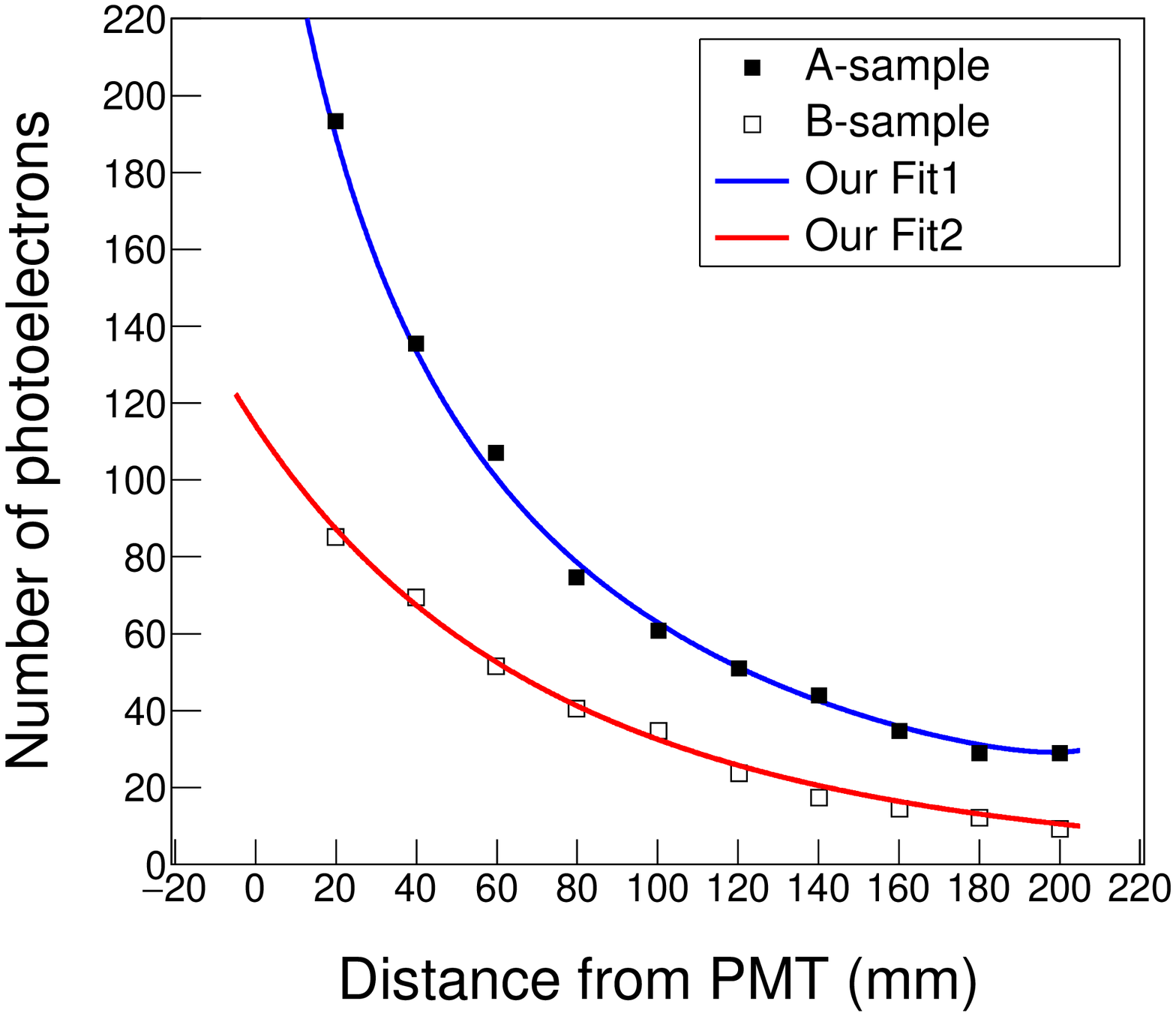}
\includegraphics[width = 0.43\textwidth]{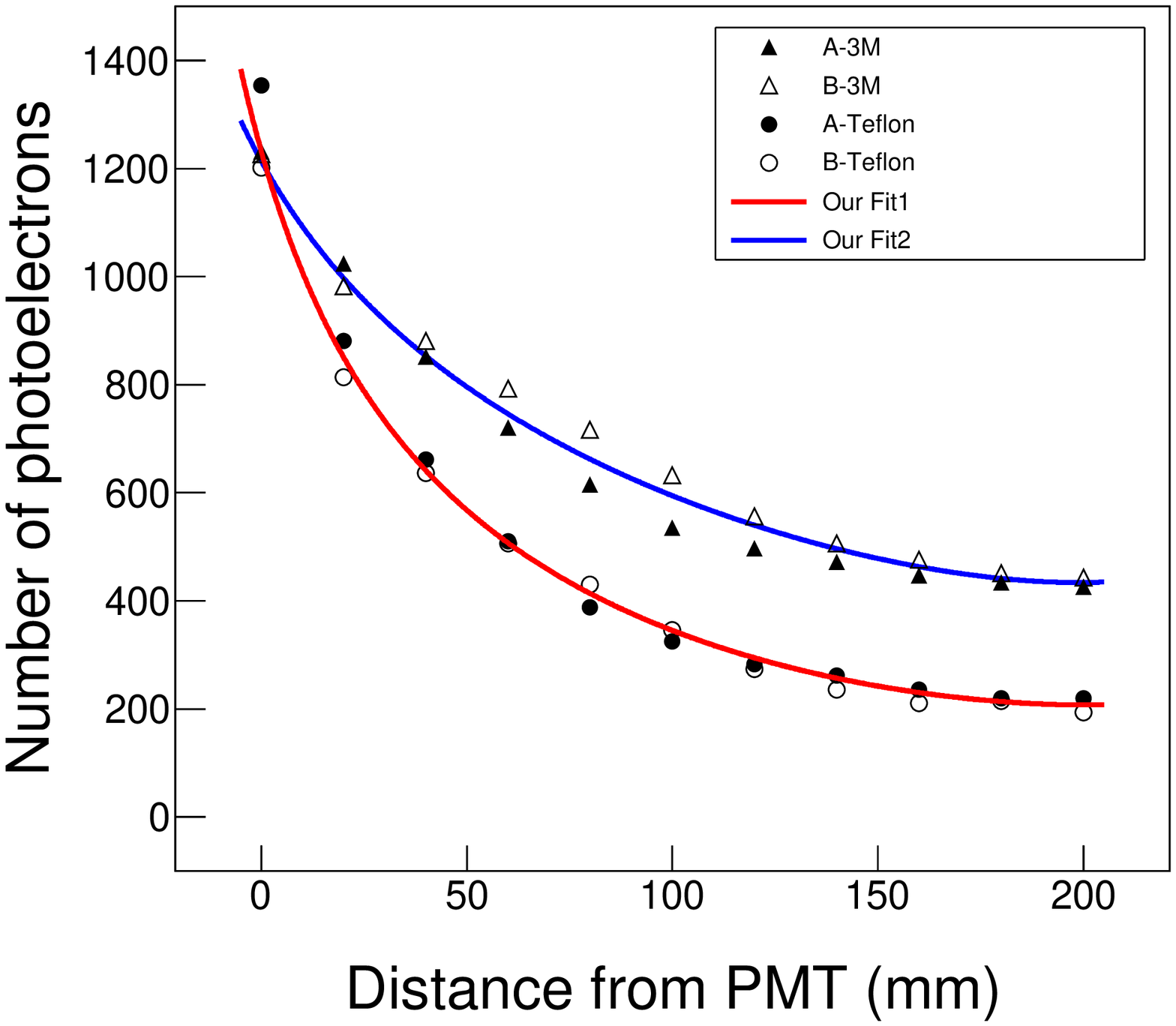}
\caption{left) Measurements with the $^{137}Cs$ source were performed for A and B plastic scintillator samples in the black box
variant; right) comparison of coating performance for the left plastic scintillator\cite{Gierlik08}. Our model fit show as line with color.}
\label{fig2.our}
\end{center}
\end{figure}

\begin{figure}
\begin{center}
\includegraphics[width = 0.73\textwidth]{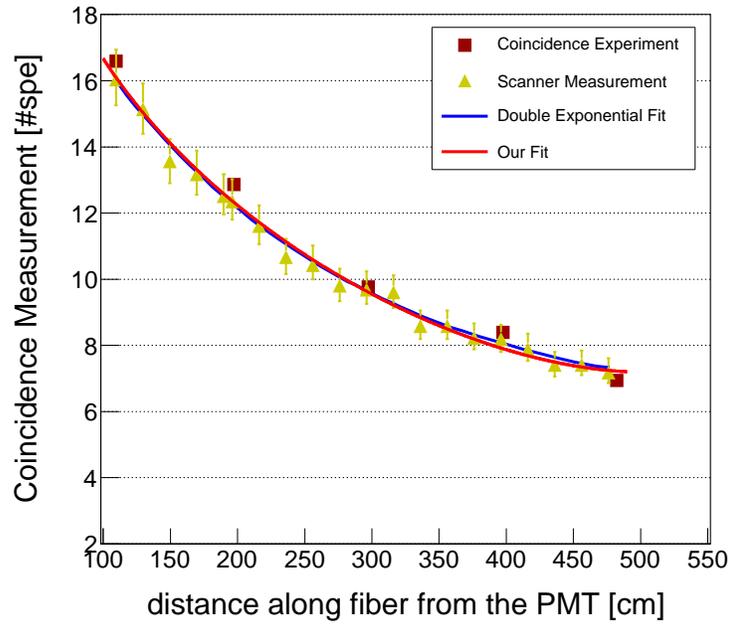}
\caption{Superposition of the number of SPEs measured with the muon telescope together with scanner measurements on the same scintillator-fiber-PMT set up. The x axis is distance along the fiber from the PMT, the y axes are signal amplitudes measured in SPE for the coincidence measurements and volts for the scanner measurements\cite{PLATINO11}. Our model fit showes as the red line.}
\label{fig3.our}
\end{center}
\end{figure}

To show how powerful our model is, we describe more experimental data from different sources,
including pilot B light output vs length using different
PMT.\cite{Kaiser64}, The effect of wrapping material on the light transmission
for different materials \cite{Taiuti96}, Light Transport in Long Plastic Scintillators\cite{Gierlik08},
and plastic scintillator muon counters used in cosmic showers detection\cite{PLATINO11}.
Fig.\ref{fig1.our}-\ref{fig3.our} show that our model fits well with all the above experimental data.
\section{Discussion}\label{Discussion}
Why is our new model so generalized? Because the spread angle is an important physical factor that describes the light propagation in the scintillator,
and this factor is considered in our new model.
Moreover, our new model includes the three traditional models from Sec.2(see Appendix for details).

Case1:when $c_1=0$ and $c_3=0$, our model can be degenerated to the ED model.

Case2:when $c_1=0$ and $c_4>>L$, our model can be degenerated to the RB model.

Case3:when $c_3=0$, $\lambda$ is about metres and larger than $L$, our model can be approximately regarded to the DE model.

There is worth to mention that a semi empirical way the reflections of light at the ends of the scintillator was developed by R.Chipaux and M.G\'el\'eoc\cite{CMS}. Our model can also be approximately regarded to this semi empirical way when $c_3=0$.
Therefore, our model can fit all experimental data that traditional models above can fit.
\section{Conclusions}\label{Conclusion}
Our model for light transport in scintillators considers the contribution of the spread angle.
It can present the data from DAMPE to a great extent, and reduces the deviation between fitting function and the sample from 29\% to less than 2\%.
In addition, it contains the above traditional models. So it is stronger than the above traditional models,
and it can fit all data that the above traditional model can fit.\\

{\bf Acknowledgments}:
This work are supported by the National Basic Research Program of China (973 Program) 2014CB845406, West Light Foundation of Chinese Academy of Sciences (Y532050XB0) and Key Research Program of Frontier Sciences, CAS, Grant NO.QYZDY-SSW-SLH006. One of us (J. Lan) thanks Z. Li, B. L. Zhang,
Z. W. Peng and Emilio CIUFFOLI for their help.
\\

{\bf Appendix}:
For case1, $c_1=0,~c_3=0$, then $f(x)=I_0e^{\frac{c_1}{c_2+x}}e^{-x/\lambda}+I_0\frac{c_3}{1-x/c_4}e^{-(2L-x)/\lambda }= I_0e^{\frac{0}{c_2+x}}e^{-x/\lambda}+I_0\frac{0}{1-x/c_4}e^{-(2L-x)/\lambda}= I_0e^{-x/\lambda}$, our model can be degenerated to the ED model.

For case2, $c_1=0,c_4>>L$, then
$f(x)=I_0e^{\frac{c_1}{c_2+x}}e^{-x/\lambda}+I_0\frac{c_3}{1-x/c_4}e^{-(2L-x)/\lambda } $
$= I_0e^{\frac{0}{c_2+x}}e^{-x/\lambda}+I_0\frac{c_3}{1-0}e^{-(2L-x)/\lambda }$
$= I_0(e^{-x/\lambda}+c_3e^{-(2L-x)/\lambda })$, our model can be degenerated to the RB model.

For case3, $c_3=0$, use $I$ to replace $I_0$,
$f(x)=Ie^{\frac{c_1}{c_2+x}}e^{-x/\lambda}+I\frac{c_3}{1-x/c_4}e^{-(2L-x)/\lambda } $
$ = Ie^{\frac{c_1}{c_2+x}}e^{-x/\lambda}+I\frac{0}{1-x/c_4}e^{-(2L-x)/\lambda }$
$=  Ie^{\frac{c_1}{c_2+x}}e^{-x/\lambda} =  I(1+\frac{c_1}{c_2+x}+\frac{1}{2!}(\frac{c_1}{c_2+x})^2+O)e^{-x/\lambda},$
with $\lambda$ being about metres and larger than $L$. For the DE model, with $\lambda_0$ being of the order of a few centimetres and  $\lambda_1$ ranging from 1 to several metres, $f_{DE}(x)=I_{0}e^{-x/\lambda_0}+I_{1}e^{-x/\lambda_1}=I(\frac{I_0}{I}e^{-x/\lambda_0+x/\lambda}+\frac{I_1}{I}e^{-x/\lambda_1+x/\lambda})e^{-x/\lambda}$
$=I(\frac{I_0}{I}e^{\frac{\lambda_0-\lambda}{\lambda\lambda_0}x}+\frac{I_1}{I}e^{\frac{\lambda_1-\lambda}{\lambda\lambda_1}x})e^{-x/\lambda}
=I(\frac{I_0}{I}e^{-\frac{1}{\lambda_0}x}+\frac{I_1}{I}e^{\frac{1}{\lambda}x})e^{-x/\lambda}$.

$f_{DE}(x)=I\{\frac{I_0}{I}[1+\frac{-x}{\lambda_0}+\frac{1}{2!}(\frac{-x}{\lambda_0})^2+O]
+\frac{I_1}{I}[1+\frac{x}{\lambda}+O]\}e^{-x/\lambda}$ for $x<\lambda_0$;

$f_{DE}(x)=I\{\frac{I_0}{I}e^{-\frac{1}{\lambda_0}x}
+\frac{I_1}{I}[1+\frac{x}{\lambda}+\frac{1}{2!}(\frac{x}{\lambda})^2+O]\}e^{-x/\lambda}$, for $ x>\lambda_0$.

Adjusting the parameters $c_1$ and $c_2$ such that $1+\frac{c_1}{c_2+x}+\frac{1}{2!}(\frac{c_1}{c_2+x})^2 \approx \frac{I_0}{I}[1+\frac{-x}{\lambda_0}+\frac{1}{2!}(\frac{-x}{\lambda_0})^2]
+\frac{I_1}{I}[1+\frac{x}{\lambda}]$ for $x<\lambda_0$, and $1+\frac{c_1}{c_2+x}+\frac{1}{2!}(\frac{c_1}{c_2+x})^2 \approx \frac{I_0}{I}e^{-\frac{1}{\lambda_0}x}+\frac{I_1}{I}[1+\frac{x}{\lambda}+\frac{1}{2!}(\frac{x}{\lambda})^2]$ for $x>\lambda_0$, our model can be approximately regarded to the DE model.\\


\end{document}